\documentclass[acmsmall]{acmart}

\AtBeginDocument{%
  }

\usepackage{longtable}
\usepackage{array}
\usepackage[normalem]{ulem}
\usepackage{xcolor}

\newcommand{\edit}[1]{\textcolor{black}{#1}}

\setcopyright{acmlicensed}
\copyrightyear{2026}
\acmYear{2026}
\acmDOI{10.1145/3831337}
\acmConference[CHI Play '26]{The Annual Symposium on Computer-Human Interaction in Play}{November 02--05,
  2018}{York, UK}
\acmISBN{978-1-4503-XXXX-X/2018/06}

\begin{document}

\title{Diagnosing Performance in Invasion-Based Esports: The Esports Performance Screening (EPS) Framework}

\author{Michael Trotter}
\authornote{Corresponding author: michael.trotter@hh.se. The authors declare no conflicts of interest.}
\orcid{0000-0003-4386-2033}
\affiliation{
\department{Department of Health and Sports}\institution{Halmstad University}
\city{Halmstad}
\country{Sweden}}
\email{michael.trotter@hh.se}

\author{Matthew Watson}
\orcid{0000-0002-0366-9821}
\affiliation{
\department{Department of Performance Psychology}\institution{German Sport University Cologne}
\city{Cologne}
\country{Germany}}
\email{matthew.watson@stud.DSHS-Koeln.de}

\author{Bastian Ils\o  Hougaard}
\orcid{0000-0002-6861-1858}
\affiliation{
\department{Department of Architecture, Design and Media Technology}\institution{Aalborg University}
\city{Aalborg}
\country{Denmark}}
\email{biho@create.aau.dk}

\author{Frauke Kubischta}
\orcid{0009-0002-7452-482X}
\affiliation{
\department{Department of Health and Sports}\institution{Halmstad University}
\city{Halmstad}
\country{Sweden}}
\email{frauke.kubischta@hh.se}

\author{Hendrik Knoche}
\orcid{0000-0003-3950-8453}
\affiliation{
\department{Department of Architecture, Design and Media Technology}\institution{Aalborg University}
\city{Aalborg}
\country{Denmark}}
\email{hk@create.aau.dk}

%
%
%
%

\begin{teaserfigure}
  \centering
  \includegraphics[width=\textwidth]{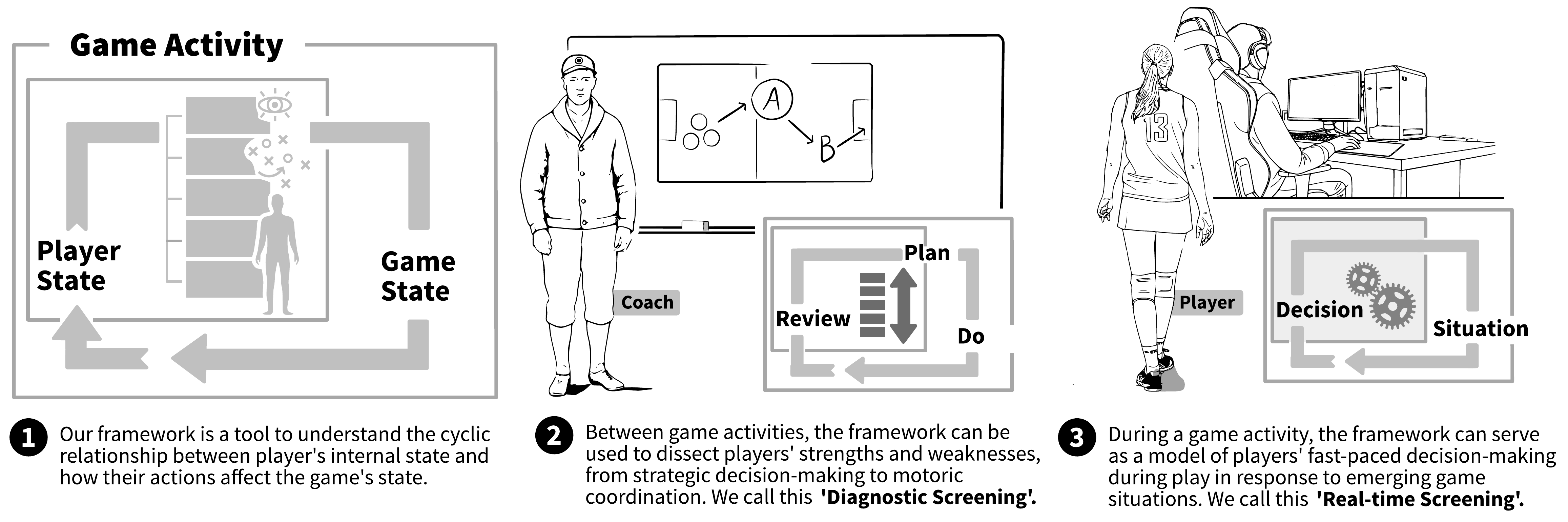}
  \Description{The figure consists of three numbered panels arranged horizontally. Panel 1 depicts “Game Activity” as a cyclical interaction between “Player State” and “Game State.” Arrows indicate continuous bidirectional influence between the internal state of the player and the evolving state of the game. Panel 2 shows a coach figure alongside a board and a “Plan–Do–Review” cycle. This panel illustrates how the framework can be used between games or training sessions to analyse strengths and weaknesses across levels from strategic decision-making to motor coordination. This process is labelled “Diagnostic Screening.” Panel 3 shows a player engaged in gameplay alongside a “Decision–Situation” loop. This represents in-game application of the framework, where players make rapid decisions in response to evolving situations. This process is labelled “Real-time Screening.” Together, the panels demonstrate that the framework operates both retrospectively (analysis and planning) and prospectively (real-time decision-making), supporting performance development across contexts.}
  \caption{Conceptual architecture of the Esports Performance Screening (EPS) framework (1) including its application in diagnostic screening (2) and real-time screening (3).
}
  \label{fig:hts}
\end{teaserfigure}
\renewcommand{\shortauthors}{Trotter et al.}

\begin{abstract}
\edit{Coaching in esports continues to become more common, yet conceptual tools for coaches to analyse performance breakdowns in esports remain limited. Existing approaches lack a way to distinguish between mental errors (i.e., mistakes relate to strategy and tactics), and physical slips (i.e., motor execution). This paper introduces the Esports Performance Screening (EPS) framework, that integrates several frameworks and models from sport and computing science which can be used to analyse mental and motor performance in esports. The EPS framework organises performance into five interconnected levels: strategy, tactics, tasks, actions and operations.} Across these levels, teams pursue forms of superiority that influence transitions between stability and instability during invasion-based esports competition. The framework supports two complementary modes of screening use: diagnostic during reflective review and real-time screening during live play. Through illustrative esports scenarios, we demonstrate how instability can originate at different hierarchical layers and how misattribution of failure can obscure underlying causes. The purpose of the EPS framework is to help coaches systematically diagnose performance breakdowns, and act as heuristics to aid gameplay analysis during a match. The paper contributes a theoretically grounded, title-agnostic esports framework to support systematic performance analysis and pedagogical development in invasion-based esports.
\end{abstract}



\maketitle

\section{Introduction}
Esports has evolved from grassroots Local Area Network (LAN) events into a global competitive industry attracting millions of spectators~\cite{newzoo_newzoo_2022}. Coaching practice in esports has not kept pace with this growth, particularly in under-resourced grassroots and educational environments~\cite{watson_i_2025,watson_introducing_2022}. Interview studies with esports coaches indicate that coaching practice is shaped by informal experience and relies on improvised, ad-hoc approaches rather than the use of structured pedagogical tools or frameworks to guide training and reflection~\cite{juntunen_challenges_2022,sabtan_current_2022,watson_introducing_2022}.
Perhaps unsurprisingly, players perceive a lack of clear reasoning and effectiveness behind the training activities conducted within their team environments~\cite{abbott_perceptions_2023}. Although video on demand (VOD) and in-game statistics feature prominently within esports training~\cite{abbott_perceptions_2023}, it appears that these digital tools are rarely used to support a broader systematic reflection on whether success or failure in playing was the result of poor decision-making or failed execution (motor or cognitive)~\cite{norman_design_2013, kleinman_trust_2024}. The purpose of the Esports Performance Screening (EPS) framework, is to provide a structured vocabulary and diagnostic logic (tool) to help coaches diagnose performance breakdowns by systematically locating whether errors originate at the strategic, tactical, task, action, or operational level of gameplay, and ultimately to promote more effective learning within esports team training environments. \edit{Theoretical application of the EPS framework is provided through hypothetical case examples based on League of Legends.}

Many of the most prominent esports titles share structural features with invasion games. Palmer et al. [13] describe invasion games as “those in which the main objectives are to invade the territory of an opponent to score and simultaneously prevent the opposition from doing so”. In traditional sport, team invasion games involve players invading opposition territory to score a goal while preventing the opponent from doing the same (Inns et al., 2023). Many esports have a very similar design, where, for example, players must enter into an opponents territory to destroy an objective (e.g., Multiplayer Online Battle Arenas (MOBA), plant an in game item on an opponent's territory (e.g., Counterstrike, Valorant) or score a goal (e.g., Rocket League, EA sports FC). These similarities between invasion games and invasion-based esports create an opportunity to apply established coaching practices from traditional invasion games to esports, with the aim of improving performance.

\edit{Performance in invasion-based esports is inherently multi-layered, including strategy, tactics and operational skills~\cite{xiaWhatContributesSuccess2019}. In traditional team invasion games, players must concurrently perform athletic skills, and team-tactical strategies, which has led researchers to adopt increasingly multi-factorial approaches to performance assessment~\cite{innsDecisionmakingAssessmentsYouth2023}. A comparable layering is evident in invasion-based esports. Players must coordinate perceptual and motor operations through digital interfaces~\cite{rogersKovaaKsAimTrainer2024}, execute game-specific actions, solve emerging tactical problems, and maintain alignment with broader team strategies~\cite{trotter2026decision}. Previous models of performance in gaming and esports describe gaming as a problem-solving process involving a problem-solving mind, inductive skills, spatial imagination, eye-hand coordination, and social competencies~\cite{nagorskyStructurePerformanceTraining2020}. Esports performance models similarly identify tactical-cognitive abilities, sensorimotor coordination, and team tactics as central performance components, while recognising that each esport has different goals, rules, affordances, and control schemes~\cite{nagorskyStructurePerformanceTraining2020}. With so many factors potentially influencing player performance, errors could originate from different possible sources such as strategic misalignment, tactical error, task-selection problem, action-level misjudgement, operational execution failure, or some interaction between these layers.}

\edit{The multi-layered nature of performance in invasion-based esports creates a diagnostic problem for coaches and analysts: the observable outcome of a play does not necessarily reveal the type of error that produced it. Norman’s distinction between mistakes and slips provides a useful vocabulary for this problem~\cite{norman_design_2013}. Mistakes occur when the player or team forms an inappropriate goal, interpretation, or plan, whereas slips occur when the intention is appropriate but execution fails}~\cite{norman_design_2013}. \edit{Recent observational work in League of Legends coaching provides an esports-specific parallel to this distinction. \citeauthor{lee_crafting_2025} found that coaches identified and discussed physical slips, such as mis-clicking or pressing the wrong key, as well as logical errors related to decision-making and game understanding (i.e., broadly corresponding to what Norman would classify as mistakes). However, physical slips were rare in the observed coaching data and were difficult for coaches to detect without external aids such as screen recordings, whereas logical errors were the most common form of coach input by both frequency and duration~\cite{lee_crafting_2025}. They further argue that esports coaching may be overly concerned with logical errors, leaving physical slips difficult to identify and correct through existing review practices~\cite{lee_crafting_2025}. This suggests that coaches require more than outcome-based reviews; they require a framework that can distinguish whether a breakdown originated in reasoning, coordination, action selection, or execution.}

Current esports training and coaching practices appear often to lack structured mechanisms for diagnosing what should be improved \cite{watson_introducing_2022}. In League of Legends, players report engaging in high volumes of play, including solo queue, scrims, and VOD review, while also learning socially by watching professional players, reviewing with peers, and emulating high-status role models~\cite{abbott_perceptions_2023}. However, this learning often occurs outside organised team environments and is shaped more by subjective norms and rank-based expectations than by individual diagnostic needs~\cite{abbott_perceptions_2023}. Players in Abbott et al.’s study described a standard pattern of training involving high volumes of games, little variety, and limited planned rest or recovery, with full games and gameplay review forming the primary means of practice~\cite{abbott_perceptions_2023}. This is problematic because some players perceived aspects of training as inefficient or difficult to make productive, particularly when solo queue and scrims did not allow repeated isolation of specific scenarios or deliberate correction of targeted weaknesses~\cite{abbott_perceptions_2023}. Coaching research reports a similar tension: League of Legends coaches often work through a review-plan-do cycle shaped by pressure to win, yet their practice occurs in a context marked by limited formal coach education, technological change, and short-term performance demands~\cite{watson_introducing_2022}. Although performance analysis, VOD review, and match statistics are increasingly used in esports, existing work also cautions that performance indicators may shift across game updates and that further research is needed to determine how such indicators can be translated into effective training strategies (Novak et al., 2020). This suggests that esports coaching does not lack appropriate data; rather, an approach to convert diagnosis and feedback into strategies for performance development at different levels of performance.

Existing theoretical and applied approaches offer useful but partial heuristics for organising and communicating analysis and feedback across different levels of esports performance. Concepts from invasion games (i.e., the tactical principles of soccer) help explain how teams create and defend space, gain superiority over, and destabilise opponents~\cite{teoldo_da_costa_tactical_2009}. In this paper, superiority refers to a game-relevant advantage that players or teams create, preserve, exploit, or deny during play, such as advantages in numbers, position, information, resources, composition, mechanical execution, or coordination. Activity-based approaches help distinguish between different levels of performance, from activity level decisions to actions and operations~\cite{bednySystemicStructuralTheoryActivity2005}. The analysis of core tasks can clarify the demands and constraints of specific game environments~\cite{hougaard_aiming_2024}. Models of stability and instability can explain how teams attempt to maintain their own organisation while disrupting the organisation of opponents~\cite{wharton_expertise_2014}. However, these approaches have not been integrated into a practical framework that allows coaches, players, and analysts to delineate between strategic, tactical, task, action, and operational levels of analysis in invasion-based esports. 

\edit{Current work on invasion games demonstrates that tactical decision-making and skill execution jointly shape performance, but instruments such as the Game Performance Assessment Instrument and related tools still treat these components as separate dimensions rather than explicitly modelling their interaction across levels of play \cite{Oslin1998GPAI, Memmert2008GPAIConceptual}. Systematic reviews of performance analysis in team invasion sports similarly show a predominance of outcome-focused indicators and separated analyses of game actions, team behaviours, and tactics, offering limited insight into how strategic intentions, tactical patterns, and individual actions are integrated in practice \citep{Lord17102020}. In esports, recent work has begun to conceptualise League of Legends as a complex team invasion sport and to develop advanced performance and win-probability models, yet these approaches largely focus on either macro strategy or individual contribution and do not connect strategic, tactical, and operational behaviours into a single, coach-facing framework \cite{Zhang2024SIDOModel,Jalovaara2024WinProbEsports}. Our framework directly addresses this gap by explicitly modelling the interdependencies between strategic plans, in-game tactical decisions, and subconscious operational behaviours in invasion-based esports, and by translating these relationships into a practical tool for diagnosis and adaptive decision-making during gameplay.}

\edit{This absence of a framework that stratifies performance across a continuum from strategic to operational levels motivates our work. This paper directly addresses this gap and makes three contributions. First, it introduces the EPS framework, an integrated conceptual model for analysing invasion-based esports across strategic, tactical, task, action, and operational levels of performance. Second, it defines superiorities as game-relevant advantages that players or teams create, preserve, exploit, or deny during play, organised into mechanical, contextual, and player superiorities. Third, it exemplifies the use of the EPS framework through theoretical illustrative League of Legends scenarios, demonstrating how the model supports diagnostic analysis—in which coaches and players identify the level at which performance strengths and weaknesses emerge—and real-time tactical analysis—in which decisions, task execution, and player actions are interpreted in relation to how teams create superiorities, maintain stability, and destabilise opponents during gameplay (see Fig. 1).}

\section{The Esports Performance Screening framework}
\begin{figure}[t]
  \centering
  \includegraphics[width=\linewidth]{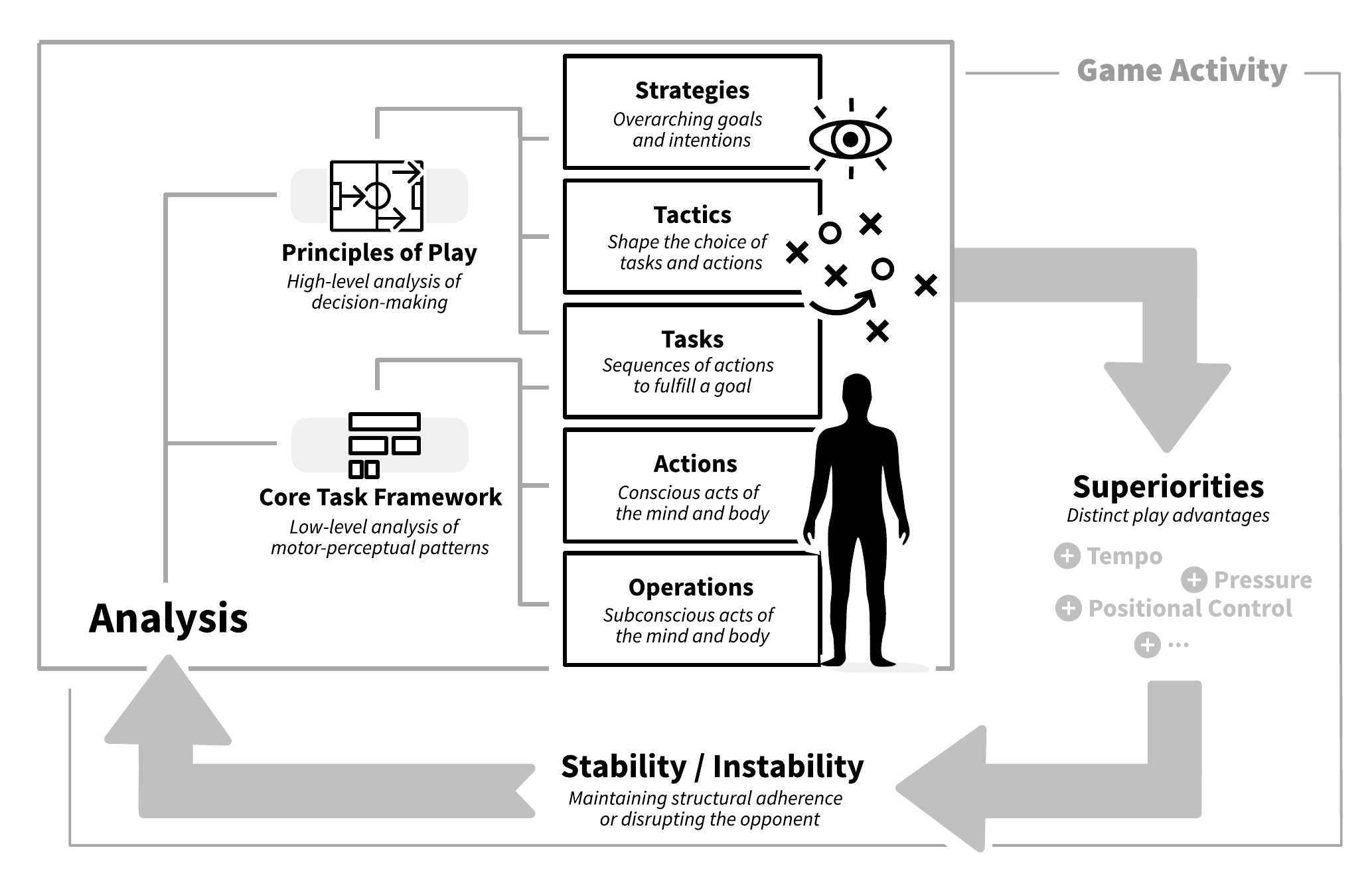}
  \Description{The figure presents a structured framework for analysing game activity across multiple levels of decision-making and action. On the left side, two analytical lenses are shown: Principles of Play (high-level analysis of decision-making) Core Task Framework (low-level analysis of motor-perceptual patterns) In the centre, a vertical hierarchy of five levels is displayed: Strategies – Overarching goals and intentions; Tactics – Shape the choice of tasks and actions; Tasks – Sequences of actions to fulfill a goal; Actions – Conscious acts of the mind and body; Operations – Subconscious acts of the mind and body. To the right, the framework connects to Game Activity, which produces Superiorities (distinct play advantages), including examples such as tempo, pressure, and positional control. Along the bottom, a horizontal arrow labeled Stability / Instability describes maintaining structural adherence or disrupting the opponent. On the far left, the term Analysis indicates that the framework is intended as a tool for systematic evaluation. The diagram shows how high-level strategic intent translates into tactical choices, concrete tasks, embodied actions, and automatic operations, which collectively shape game state and produce competitive advantages.}
  
  \caption{\textbf{The Esports Performance Screening framework.}Performance is conceptualised as a cyclical interaction between player state and game state. The framework integrates five hierarchical levels (strategy, tactics, tasks, actions and operations) and two temporal modes: diagnostic screening (retrospective traversal) and real-time screening (live regulation under time pressure).}
  \label{fig:integratedFramework}
  \Description{Integrated framework for analysing invasion-based esports.}
\end{figure}

The EPS framework (see Fig.~\ref{fig:integratedFramework}) is designed to analyse performance across multiple interconnected levels of decision-making and execution while accounting for how these processes unfold over time. Its primary purpose is diagnostic: to enable practitioners to locate the most critical errors in performance within a stratified hierarchy and to distinguish between slips of execution and mistakes of reasoning. Fig.~\ref{fig:hts} illustrates the conceptual architecture of the EPS framework. At its core, performance is understood as a dynamic interaction between player state and game state (Panel 1). Decisions at different hierarchical levels alter the competitive system, and the evolving game state in turn constrains subsequent decisions.

\subsection{Framework Selection and Construction}

\edit{The EPS framework, was developed as a conceptual synthesis rather than as an empirically validated model. Its purpose is to integrate existing theoretical frameworks and models that explain different but complementary layers of performance in invasion-based esports. Specifically, the model draws on four bodies of work: the Principles of Play from soccer~\cite{teoldo_da_costa_tactical_2009}, the Core Tasks Framework from video gaming~\cite{hougaard_aiming_2024}, the Stability/Instability Exchange Model from interceptive sports coaching~\cite{wharton_expertise_2014}, and a set of superiority categories derived from sport and esports performance analysis.}

\edit{The Principles of Play were included because they describe the strategic and tactical logic of invasion games. In soccer, principles of play provide general organising ideas that guide team behaviour in attack and defence, such as creating space, progressing toward the target, maintaining defensive compactness, and preventing the opponent from advancing. \citeauthor{teoldo_da_costa_tactical_2009} frame principles of play as a way of analysing tactical behaviour in relation to the functional demands of the game \cite{teoldo_da_costa_tactical_2009}. In the EPS framework, the Principles of Play therefore are connected to the strategy and tactics levels of the framework: they describe the broad strategic and tactical intentions that teams pursue during gameplay. }

\edit{The Core Tasks Framework was included because it "outlines the basic motor and perceptual tasks that games require in order to interact with game mechanics"~\cite{hougaard_aiming_2024}. Coaching in esports has been overly concerned with logical errors~\cite{lee_crafting_2025}, yet a player's mastery of the specific in game mechanics are also an important aspect of performance~\cite{sharpeIndexingEsportPerformance2023}. In the EPS framework, the Core Tasks Framework describes the basic motor and perceptual tasks  through which strategic and tactical intentions are enacted. }

\edit{The superiority categories were created and included to make the framework analytically usable. The superiority categories were developed through iterative discussion among the authors. Drawing on the authors’ collective experience in esports, coaching, performance analysis, and game research. Candidate superiorities were first identified from common patterns of advantage across invasion-based esports. These categories were then grouped into three higher-level types: mechanical, contextual, and player superiorities. Disagreements about category boundaries were resolved through discussion, with attention to whether each proposed superiority could be clearly defined, whether it could plausibly be gained or lost during play, and whether it contributed meaningfully to the stability or instability of the game state. The resulting taxonomy was then stress-tested against common situations from several invasion-based esports, including; bomb-site retakes in Counter-Strike 2, offensive actions in StarCraft, and ganking or objective contests in League of Legends.  This process was used to examine whether each superiority category could describe meaningful advantages across different titles and whether the categories could account for both successful and unsuccessful transitions in play.}

\edit{The Stability/Instability Exchange Model was included because it provides a way to conceptualise how coaches interpret dynamic changes in advantage during opposed play. Wharton’s model was developed from expert coaches in interceptive sports and is intended to explain how coaches filter, analyse, and comprehend environmental information during complex, unfolding action~\cite{wharton_expertise_2014}. The model proposes that expert coaches perceive play as a reciprocal interaction of opposing and affiliated forces, using conceptualisations and personalised analogies to identify the most relevant streams of information. In practical terms, this means that coaches do not only observe isolated actions; they interpret whether attacking and defensive structures are becoming more stable or unstable. Wharton further argues that scoring opportunities are often crafted rather than passively received, as teams manipulate an opponent’s structure to create signs of instability in their defensive organisation~\cite{wharton_expertise_2014}. In the EPS framework, the Stability/Instability Exchange Model therefore provides the dynamic diagnostic layer: it explains how coaches and analysts may interpret changes in the rapport of strength between opposing teams, based on how the teams strategies, tactics and behaviours generate superiorities.}

The EPS framework operates temporally in two complementary modes. First, it supports Diagnostic Screening (Fig.~\ref{fig:hts}, panel~2), a retrospective traversal of the hierarchy conducted during performance review. This allows coaches and players to systematically identify whether instability emerged from strategic misalignment, inappropriate tactical principle selection, task selection, flawed action-level judgement, or operational execution failure. Second, it supports Real-Time Screening (Fig.~\ref{fig:hts}, panel~3), in which players regulate behaviour during live game-play by anchoring their coordination of behaviour to shared tactical principles under time pressure. In both cases, the same hierarchical structure governs analysis and regulation. Across all levels, teams pursue forms of superiority that influence whether the competitive system tends toward stability or instability. By formally articulating how hierarchical decisions and evolving game states interact over time, the EPS framework provides a structured vocabulary and diagnostic logic that bridges reflective review and in-game cognition.

\subsection{Levels of the EPS framework for Analysing Invasion-Based Esports}

At every level of the framework, teams pursue some form of superiority that improves their own position or constrains the opponent. At the operations level, actions are realised through automated perceptual and motor routines whose precision and efficiency can be examined using the CTF~\cite{hougaard_aiming_2024}. At the action level, players make discrete behavioural selections and situational adaptations in response to unfolding information. At the task level, these actions are organised around an immediate objective that must be achieved under current game conditions, specifying what must be accomplished to preserve or generate superiority. At the tactical level, players develop or adapt situation-specific plans that generate localised advantages, for example forcing rotations to gain a numerical superiority or gaining a positional superiority by selecting specific avatar or unit positioning. At the strategic level, teams establish long-term advantages that shape the conditions of play, such as how they will prioritise unit or compositional scaling or control over key areas of the map. The pursuit of superiorities thus offers a unifying principle across the hierarchy and prepares the ground for the subsequent analysis of stability and instability in competitive play.

\subsection{Integrating the Principles of Play and the Core Task Framework Across Framework Levels}
In the invasion games framework the Principles of Play~\cite{clemente_systemic_2014} provide higher-order organising guidelines for collective behaviour, helping teams coordinate action in relation to teammates, opponents, space, time and the ball. Within the EPS framework, PoP therefore inform the strategy and tactics levels by describing the shared intentions that guide how teams organise and adapt during play. By contrast, the Core Task Framework clarifies how these intentions are enacted through the specific motor tasks which are preformed during the game. The CTF~\cite{hougaard_aiming_2024} distinguishes the motor demands involved in gameplay, such as aiming, pointing, steering and activation (see Table~\ref{tab:tasks-in-games}), allowing coaches to examine whether breakdowns arise from action selection, perceptual judgement, or operational execution. The EPS framework connects these components so that tactical intentions can be traced downward into specific task, action and operation demands, while mechanical limitations can also be traced upward to understand how they constrain viable tactical choices.

The following sections first unpack the diagnostic hierarchy of the EPS framework by defining each level of analysis in turn: operations, actions, tasks, tactics, and strategy. This diagnostic account explains how coaches and analysts can locate performance breakdowns across the hierarchy, from failures of execution through to failures of strategic alignment. After establishing these hierarchical components, the paper then turns to the temporal dimension of the framework, explaining how the same levels can inform real-time screening during live play when decisions must be made under time pressure.

\section{Diagnostic screening}

\subsection{Operations}

The operations level represents the most granular level of esports performance, which consists of the automated motor and perceptual processes that underlie specific actions. According to Hougaard \& Knoche, operations are the unconscious methods used to perform actions, which are determined by immediate conditions rather than a conscious goal~\cite{hougaard_aiming_2024}. In the context of esports, a player does not consciously decide to displace a mouse a specific distance or depress a specific key. Instead, these are the automated motor sub-routines that are triggered to fulfill a higher-order intent. Errors at the operations level are distinct from errors in decision-making as they represent a failure of execution rather than a failure of logic. To operationalise this layer, we drew on mechanical inputs in the CTF, specifically~Activation, which Hougaard and Knoche define as the input-based triggering of game functions~\cite{hougaard_aiming_2024}. ~\citet{lee_crafting_2025} classify errors at this level as "physical slips" where a player understands the correct course of action but fails to physically input the command~\cite{lee_crafting_2025}. Their observational study found that elite coaches frequently overlook these operational errors in favour of strategic feedback because they lack the diagnostic frameworks to identify micro-mechanical failures. In \citeauthor{norman_design_2013}'s terms~\cite{norman_design_2013}, these are slips: the intention and plan are appropriate, but the execution falters. By contrast, when the motor sequence is carried out correctly, but the underlying goal or tactical plan is inappropriate, the error reflects what Norman defines as a mistake. Therefore, recognising operations as a distinct level allows practitioners to distinguish between a "bad play" caused by a lack of tactical knowledge and one caused due to insufficient skill.

\subsection{Actions}
The action level represents the conscious implementation of tactical decisions through complex goal-directed units that combine mental prediction with motor execution. In the CTF, an action is not merely a physical movement; it is a discrete functional unit composed of both~mental operations~(perceiving and calculating) and~motor operations~(executing the movement)~\cite{hougaard_aiming_2024}. For example, the action of "landing a skill-shot" requires the player to first mentally calculate the target's velocity and trajectory (Mental Operation) and then coordinate their hand to execute the specific input (Motor Operation). This level is distinct from the operations level because it requires conscious feedback loops; the player actively monitors the cursor's position relative to the target and adjusts in real-time. We adapt the CTF~\cite{hougaard_aiming_2024}, which identifies six distinct motor tasks. While the full framework includes~``Drawing''~and~``Typing'', these are peripheral to the primary mechanics of invasion-based esports. Therefore, the EPS framework focuses on the four interaction types central to character control and combat: Pointing\textbf{:}~Requires the player to continuously align a cursor with a target with constant visual feedback (e.g., selecting a unit). Aiming\textbf{:}~Imposes a higher cognitive load as the player must predict a ballistic trajectory and collision point~\emph{without}~immediate outcome signification (e.g., a projectile taking time to travel). Steering\textbf{:}~Involves the continuous mental projection of a path to navigate an avatar through dynamic obstacles. Activation\textbf{:}~The deliberate triggering of a function (distinct from the unconscious operation of the button press) often coupled with aiming or pointing. \citet{hougaard_aiming_2024} highlight that "Aiming" specifically requires the user to internalise the physics of the game world (e.g., projectile speed), making it a distinct cognitive-motor skill from simple "Pointing." By distinguishing between the mental prediction and the motor execution within these specific actions, coaches can diagnose whether a player failed because they misread the game physics (an Action-level cognitive error) or because they lacked the motor skills to effectively execute the action or task.

\subsection{Tasks}

The Task level represents a structured unit of gameplay organised around a specific task goal and realised through coordinated actions. Within the CTF, a task is defined as a logical division of work within an activity, composed of one or more actions directed toward achieving an objective under particular conditions~\cite{hougaard_aiming_2024}. Unlike individual actions, which are discrete behavioural units, tasks organise sequences of actions toward the accomplishment of a concrete situational objective. In the EPS framework tasks are inherently conditional, they are shaped by environmental constraints and opponent behaviour. In invasion games, these conditions are dynamic and adversarial, requiring tasks to adapt continuously to shifting patterns of stability and instability~\cite{palmer_cooperative_2023}. For example, securing a contested objective, stabilising after losing numerical superiority, or converting temporary positional control into a durable advantage each represent task-level objectives. These are not isolated, but multiple coordinated actions directed toward a shared outcome~\cite{hougaard_aiming_2024}. The task level therefore structures gameplay into coherent, goal-directed segments that can be analysed and evaluated independently of the specific tactical principles that later regulate their coordination.

\begin{table}[th]
\caption{\edit{Adapted excerpt of \citeauthor{hougaard_aiming_2024}'s core task list~\citep{hougaard_aiming_2024}, showing the four most common motor tasks in invasion games, with examples of their application in League of Legends.
}}
\label{tab:tasks-in-games}
\renewcommand{\arraystretch}{1.4}
\centering
\resizebox{\linewidth}{!}{
\begin{tabular}{p{2.0cm}p{1.5cm}p{5.8cm}p{6.85cm}}
\toprule
Task & \edit{Label} & \edit{Definition} & \edit{Task and Goal League of Legends Example} \\
\hline
\raisebox{-.7\height}{\includegraphics[width=\linewidth]{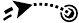}} & Aiming \newline \textit{motor} & \edit{``Accurately pointing at a target (possibly using a device) and/or predicting the collision between two objects, without outcome signification.''} & \edit{Successfully aiming and landing an AoE attack at an opponent (task) to secure local numerical superiority in that area (goal).} 
\\
\raisebox{-.7\height}{\includegraphics[width=\linewidth]{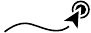}} & \edit{Pointing} \newline \edit{\textit{motor}} & \edit{``Accurately pointing at an accessible target with feedback about current pointing position.'' }& \edit{Swiftly clicking an opponent while they are still within range of a direct attack (task), to apply pressure (goal).}  \\
\raisebox{-.9\height}{\includegraphics[width=\linewidth]{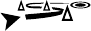}} & \edit{Steering} \newline \edit{\textit{motor}} & \edit{``Moving or guiding an object along a trajectory.''}
& \edit{Steering through the jungle stealthily (task), to secure the dragon objective (goal).} \\
\raisebox{-.7\height}{\includegraphics[width=\linewidth]{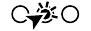}}  & Activation \newline \textit{motor} & \edit{``Initiating another mechanical system, function, or item.''} &  \edit{Activating abilities in correct order and timing (task), to maximize damage per second and increase tempo (goal).} \\
\centering ... & \centering ... & \centering ... & \centering ...
\vspace{0.25cm}
\end{tabular}}
\end{table}

\subsection{Tactics}

The Tactical level governs the selection and sequencing of the PoP that regulate collective organisation during live gameplay. Drawing on the core tactical principles articulated by~\citet{teoldo_da_costa_tactical_2009} and generalised in Table~\ref{tab:PoP}, this level provides the organising logic through which teams coordinate their offensive and defensive structures and collective movement in response to evolving game states. Whereas the task level specifies what must be achieved under current conditions, the tactical level determines the collective principles guiding how that objective is pursued. Tactical regulation operates under constrained time horizons and high contextual specificity~\cite{buekers_strategy_2020}. In invasion-games opponents offensive and defensive offensive decisions and behaviours can rapidly alter the balance of stability and instability. At this level, the core PoP function as structured heuristics that reduce decision complexity by anchoring collective behaviour to a coherent organising logic. For example, selecting a defensive principle such as Delay or Balance shifts emphasis toward time generation and spatial containment, whereas selecting an offensive principle such as Penetration or Offensive Unity shifts emphasis toward structured advancement and coordinated pressure.

\begin{table}[htbp]
\caption{Core principles of play linking defence-attack  and/or attack-defence transitions adapted from~\citeauthor{teoldo_da_costa_tactical_2009}'s core tactical principles of play. We generalised their terminology to enhance applicability across invasion games.}
\label{tab:PoP}
\centering
\renewcommand{\arraystretch}{1.7}
\begin{tabular}{p{0.47\textwidth}p{0.47\textwidth}}
\hline
\textbf{Attack} & \textbf{Defence} \\
\hline

\textbf{Penetration} \newline
\small
-- Destabilise the opponent’s defensive organisation; \newline
-- Directly attack an opposing player or the primary objective; \newline
-- Create advantageous attacking situations in numerical and \newline \hphantom{-- }spatial terms.
&
\textbf{Delay} \newline
\small
-- Decrease the space the initiating player has for offensive \newline \hphantom{-- }action; \newline
\-- Direct the progression of the initiating player; \newline
-- Block or delay the opponent's attack or counter-attack; \newline
-- Provide more time for defensive organisation; \newline
-- Restrict progression options to other opponents; \newline
-- Prevent direct advancement into critical defensive space; \newline
-- Prevent immediate execution of the primary objective.
\\
\hline
\textbf{Offensive Coverage} \newline
\small
-- Support the initiating player by providing options to maintain \newline \hphantom{-- }continuity of play; \newline
-- Decrease opponents' pressure on the initiating player; \newline
-- Create numerical superiority; \newline
-- Unbalance the opponent's defensive organisation; \newline
-- Ensure conservation of controlled resources or positional \newline \hphantom{-- }advantage.
&
\textbf{Defensive Coverage} \newline
\small
-- Act as a secondary obstacle to the initiating player in case \newline \hphantom{-- }delay is bypassed; \newline
-- Provide support to the player performing delay to reinforce \newline \hphantom{-- }blocking initiative.
\\
\hline
\textbf{Width and Length} \newline
\small
-- Use and enlarge the effective play-space of the team; \newline
-- Expand distances between opponents' positions; \newline
-- Make tracking or direct containment more difficult; \newline
-- Facilitate offensive actions; \newline
-- Move to a safer space; \newline
-- Win time for a better decision; \newline
-- Seek safe options through defensive-positioned players \newline \hphantom{-- }to continue play.
&
\textbf{Balance} \newline
\small
-- Ensure defensive stability in the area of primary contest; \newline
-- Support teammates performing Delay and Defensive\newline \hphantom{-- }Coverage; \newline
-- Block progression options; \newline
-- Monitor and contain advancing opponents; \newline
-- Pressure the initiating player; \newline
-- Re-establish control and relocate play away from objectives.
\\
\hline

\textbf{Depth Mobility} \newline
\small
-- Disrupt opponent's defensive organisation; \newline
-- Position suitably to pursue the primary objective; \newline
-- Create in-depth movement options; \newline
-- Achieve controlled access to contested space.
&
\textbf{Concentration} \newline
\small
-- Increase protection of the primary defensive objective; \newline
-- Direct offensive actions toward lower-risk areas; \newline
-- Increase defensive pressure in high-threat zones.
\\
\end{tabular}
\end{table}

\begin{table}[htbp]
\renewcommand{\arraystretch}{1.7}
\begin{tabular}{p{0.45\textwidth}p{0.50\textwidth}}
\hline
\textbf{Offensive Unity} \newline
\small
-- Facilitate coordinated advancement; \newline
-- Attack in unity; \newline
-- Increase structural security; \newline
-- Enable coordinated involvement of multiple players; \newline
-- Maintain relational cohesion.
&
\textbf{Defensive Unity} \newline
\small
-- Enable the team to defend in unity; \newline
-- Ensure spatial stability and dynamic synchronisation; \newline
-- Decrease the offensive amplitude of the opponent in width \newline \hphantom{-- }and depth; \newline
-- Establish guiding structural references outside the primary \newline \hphantom{-- }contest zone; \newline
-- Continuously rebalance defensive organisation; \newline
-- Obstruct progression options within contested areas; \newline
-- Enable subsequent defensive actions; \newline
-- Increase collective presence within high-threat zones.
\\
\hline
\end{tabular}
\end{table}

\subsection{Strategy}

The strategic level represents the highest-order dimension of performance, defining the team’s overarching intentions and goals for the session, whether competitive or developmental. In the context of the CTF, this level directs how players approach completing  the activity, defined by  \citet{hougaard_aiming_2024} as a coherent system of processes directed toward a specific~motive. For the purposes of the EPS framework, we adopt the term strategy for the highest level of analysis, possible within a game activity in the CTF. We do so not only because strategy is more immediately recognisable in applied esports coaching, but also because it better captures the temporally extended and multi-layered nature of this level. Strategic concerns range from long-term organisational and player-development goals, through season- and tournament-level planning, to match-specific approaches to particular opponents. Although this paper focuses on match and training contexts, we acknowledge that the strategic level in any given moment is shaped by these broader constraints and objectives.

In esports, the motive shapes the strategy and for example, in a competitive match, the motive is victory, and the strategy defines the "win condition" (e.g., controlling map objectives). However, in grassroots or educational settings, the motive may be skill acquisition. Here, the strategy serves a pedagogical function (e.g., a coach instructing a team to play aggressively to practice engaging, regardless of the match outcome). Strategic decisions thus establish the operational boundaries, the 'rules of engagement' for the session, guiding which tactics are permissible and which are counter-productive to the overarching goal. This distinction aligns with decision-making literature in sport, where expert performance is characterised by the ability to form sophisticated mental models prior to engagement~\cite{abraham_taking_2011}. These models act as a scaffold. For a professional team, the scaffold ensures all players share a vision for winning~\cite{TrotterGoals2025}. For a novice team, the scaffold directs attention to specific learning outcomes during training. For instance, in a scrim, a team might adopt a strategy of "fighting for every dragon" to test their limits. If they lose the game but successfully execute the fights, the "Activity" (development) was successful, even if the "Match" was lost. This highlights that strategy is not a fixed pursuit of victory, but a flexible framework adapted to the specific context of play. Consequently, the strategic level provides the necessary continuity across a match, ensuring that tactical adaptations and motor actions remain aligned with the team's ultimate objective; whether that objective is related to competition or training.

\section{Real-time Screening}

\subsection{Superiority in Esports}
Superiorities are defined here as the game-relevant advantages that players or teams create, preserve, exploit, or deny during play (Table 2). Within the EPS framework, superiorities provide the link between hierarchical decision-making and the evolving state of the game. Strategic, tactical, task, action, and operational decisions are not made in isolation; they are enacted to improve a team’s own position, constrain the opponent’s options, or shift the game state toward greater stability for one team and greater instability for the other.

The concept of superiority in existing literature tends to discuss specific forms of advantage (i.e., numerical superiority~\cite{rumpfManipulatingPlayerNumbers2026, vilarVaryingNumbersPlayers2014}), rather than providing a unified taxonomy of the different ways in which superiority may be created during play. However, numerical superiority alone does not adequately capture the range of advantages that matter in esports. In theory a team may lose a fight despite having more players present if the opponent has stronger items, better positional control, superior information, greater coordination, or a more favourable composition. Conversely, a team may overcome a numerical disadvantage if it controls terrain, has better timing, possesses more accurate information, or executes a coordinated transition more effectively.

\begin{table}[t]
\caption{\parbox{\dimexpr\linewidth-4em\relax}{Definitions of types and sub-types of superiorities in esports}}
\centering
\hyphenpenalty=10000
\exhyphenpenalty=10000
\emergencystretch=2em
\label{tab:sup}
\centering
\small
\begin{tabular}{p{0.10\textwidth} p{0.14\textwidth} p{0.7\textwidth}}
\hline
\textbf{Superiority} & \textbf{Subtype} & \textbf{Definition} \\
\hline

Mechanical & Quantitative& Possessing upgraded or higher-tier units, abilities, or equipment that confer a direct mechanical advantage. \\

& Informational & Possessing more accurate or timely knowledge of the opponent’s location, composition, or intentions. \\

& Economic & Access to greater in-game resources, such as gold, fuel, or currency. These resources translate into material advantages (i.e., more powerful units, earlier upgrades, or stronger loadouts). \\

\hline

Contextual & Compositional & The effective arrangement of avatars, champions, classes, or units relative to the opponent’s lineup (synergies, counter-matchups, or role distribution that favour one side). \\

& Numerical & Involves having more controllable entities or players engaged in a given encounter. \\

& Positional & Control over space or terrain that offers tactical benefits (i.e., cover, high ground, flanking routes). \\

\hline

Player & Perceptual & Capacity to process and act on multiple streams of in-game information simultaneously (i.e., situational awareness). \\

& Qualitative & Differences in skill related to mechanical execution (i.e., actions per minute), tactical decision-making, and game sense \& game knowledge. \\

& Coordination & The ability to coordinate and execute strategies more effectively. This can also include surprising and deceiving the opponent, as well as understanding when to transition from offense to defense. \\

\hline
\end{tabular}
\end{table}

For this reason, the EPS framework distinguishes three broad categories of superiority: mechanical, contextual, and player superiority (see table 3). Mechanical superiorities refer to advantages embedded in the game system itself, including technical, informational, and economic advantages. These include stronger equipment or abilities, access to more accurate or timely information, and greater in-game resources such as gold, fuel, or currency. Contextual superiorities refer to advantages generated by the current configuration of the game state, including compositional, numerical, and positional advantages. These describe how avatars, units, players, and spatial relations create favourable or unfavourable conditions for action. Player superiorities refer to advantages arising from the capabilities of the players or team, including perceptual capacity, qualitative skill differences, and coordination. These describe how well players perceive, decide, execute, communicate, and synchronise their actions under pressure.

These categories are intended to be analytical rather than exhaustive or mutually exclusive. In practice, superiorities often interact. For example, gaining economic superiority may lead to technical superiority through upgraded items, which may then support positional superiority by allowing a team to contest space more safely. Similarly, informational superiority may enable better coordination, which may then create numerical superiority at a local point of contest. The purpose of the taxonomy is therefore not to assign every event to a single fixed category, but to help coaches and analysts describe how advantages emerge, combine, and decay over time.

In the EPS framework, superiorities influence the stability or instability of the game state. When a team holds sufficient superiorities to pursue its intended plan, the game state is relatively stable for that team. When key superiorities are lost, denied, or misused, the team may be forced into instability, requiring tactical adaptation, retreat, reset, or recovery. The analysis of superiorities therefore helps explain not only whether a team is advantaged, but why that advantage exists, where it is located in the hierarchy, and how it may be preserved or disrupted.

\subsection{The Stability--Instability Cycle}

\citeauthor{wharton_expertise_2014}'s Stability and Instability Exchange Model (see Fig. 3) describes play as an ongoing contest over how ``stable'' each team's situation is, and it cycles through two phases: offence and defence. In the offensive phase, a team attempts to create instability in the opponent's defensive organisation by applying tactics that remove resources or reduce the opponent's ability to respond (e.g., damaging key structures, forcing rotations, or eliminating players or units)~\cite{wharton_expertise_2014}. In the defensive phase, the opponent attempts to absorb or blunt the attack so that the attacker's plan becomes unstable and can no longer be sustained, forcing a reset, a retreat, or a turnover. In this model, instability is the state where a team's current plan cannot be sustained without unacceptable risk, because the opponent's actions have degraded resources, positioning, timing, or coordination. When instability occurs in either an offensive or defensive phase, teams have two possible options; collapse (i.e., losing decisively), or retreat/reset, where the team concedes the immediate contest by for instance disengaging, giving up space or an objective to prevent collapse. Stability is the inverse: a state in which the team retains sufficient resources, positioning, timing, and coordination to enact its intended strategy and tactics with effective agency. A useful analogy is to think of stability like a chair. Each loss of resources, position, timing, or coordination is like losing a leg: the chair may still stand, but it becomes easier to tip. As legs are lost, the position becomes increasingly fragile, and one additional error can trigger collapse (for example, conceding an objective, losing map control, or suffering a decisive wipe).

\begin{figure}[t]
  \centering
  \includegraphics[width=0.55\linewidth]{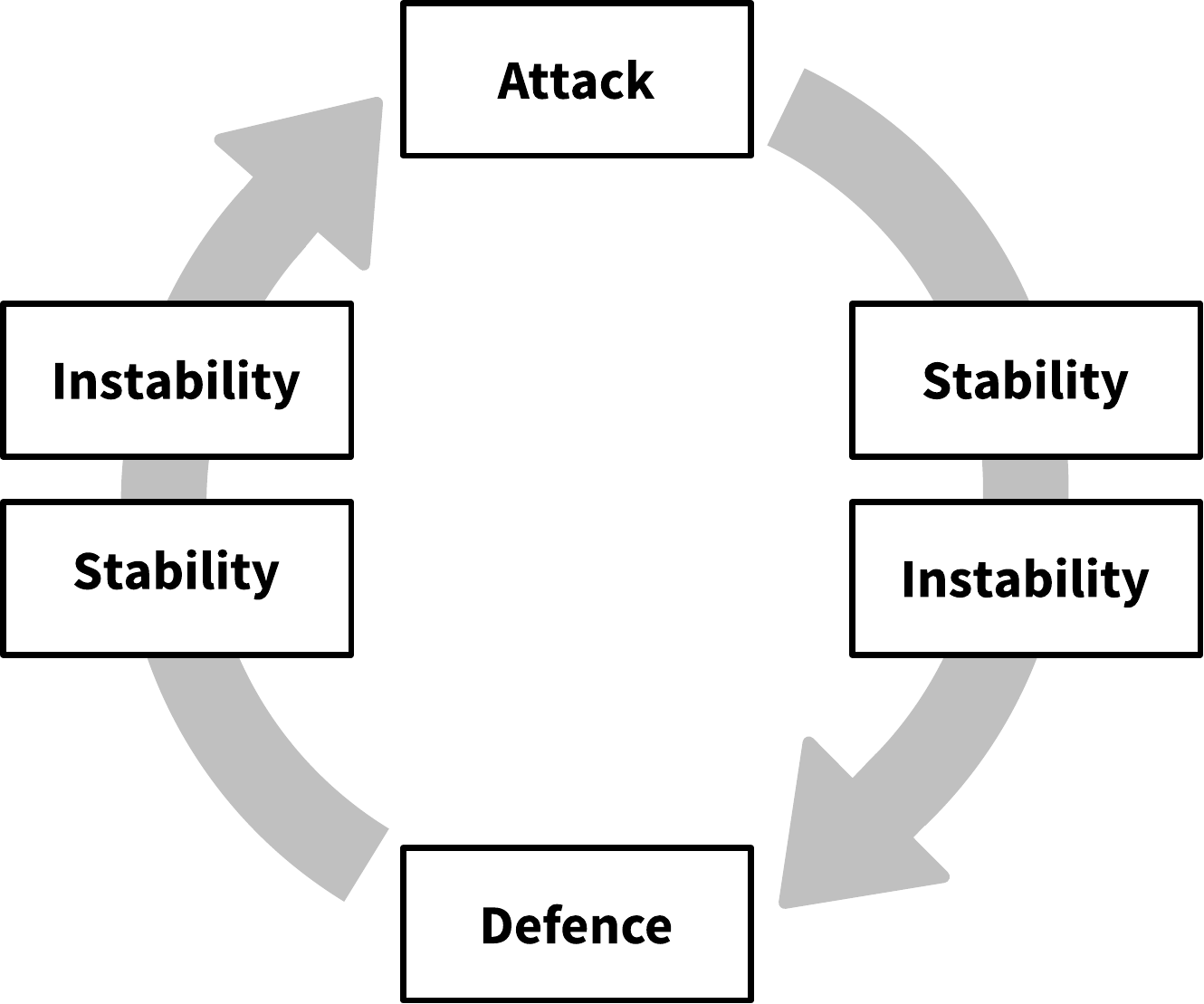}
  \Description{The figure presents a circular flow diagram with four main labels positioned around a clockwise arrow. At the top is “Attack,” at the bottom is “Defence.” On the right side, the sequence reads “Stability” above “Instability.” On the left side, the sequence reads “Instability” above “Stability.” The diagram illustrates that during attack, a team moves from stability to instability in order to disrupt the opponent. Conversely, during defence, the team seeks to regain stability after instability. The circular arrow indicates that this is a continuous, cyclical process, with teams constantly transitioning between stable and unstable states as possession changes.}
  \caption{\edit{\textbf{The Stability–Instability Exchange Cycle.} Adapted from \citet{wharton_expertise_2014}. Teams alternate between offensive and defensive attempts to induce instability. Instability represents a state in which resources, positioning, timing or coordination are insufficient to sustain the current plan without unacceptable risk.}}
  \label{fig:Stability–Instability}
  \Description{Integrated framework for analysing invasion-based esports.}
\end{figure}

\section{The Framework in Practice: Diagnostic and Real-time Screening}

During diagnostic screening, such as during post-game video review, the EPS framework is viewed vertically, offering a diagnostic path up and down the hierarchy (i.e., strategy, tactics, tasks, actions and operations) to pinpoint which layer would produce the most effective improvements to performance. Diagnostic screening can only occur in situations where decision-making is not constrained by time~\cite{buekers_strategy_2020}. Conversely, real-time screening in which analysis occurs during live game-play, where decisions are heavily constrained by time pressure. As a result of time pressure players are unlikely to consciously traverse the full hierarchy. In practice, real-time screening operates primarily at the tactical, task and action levels: players use PoP as heuristics to make rapid judgements about the unfolding situation and to select and implement an appropriate tactic.  

\begin{figure}[t]
  \centering
  \includegraphics[width=\textwidth]{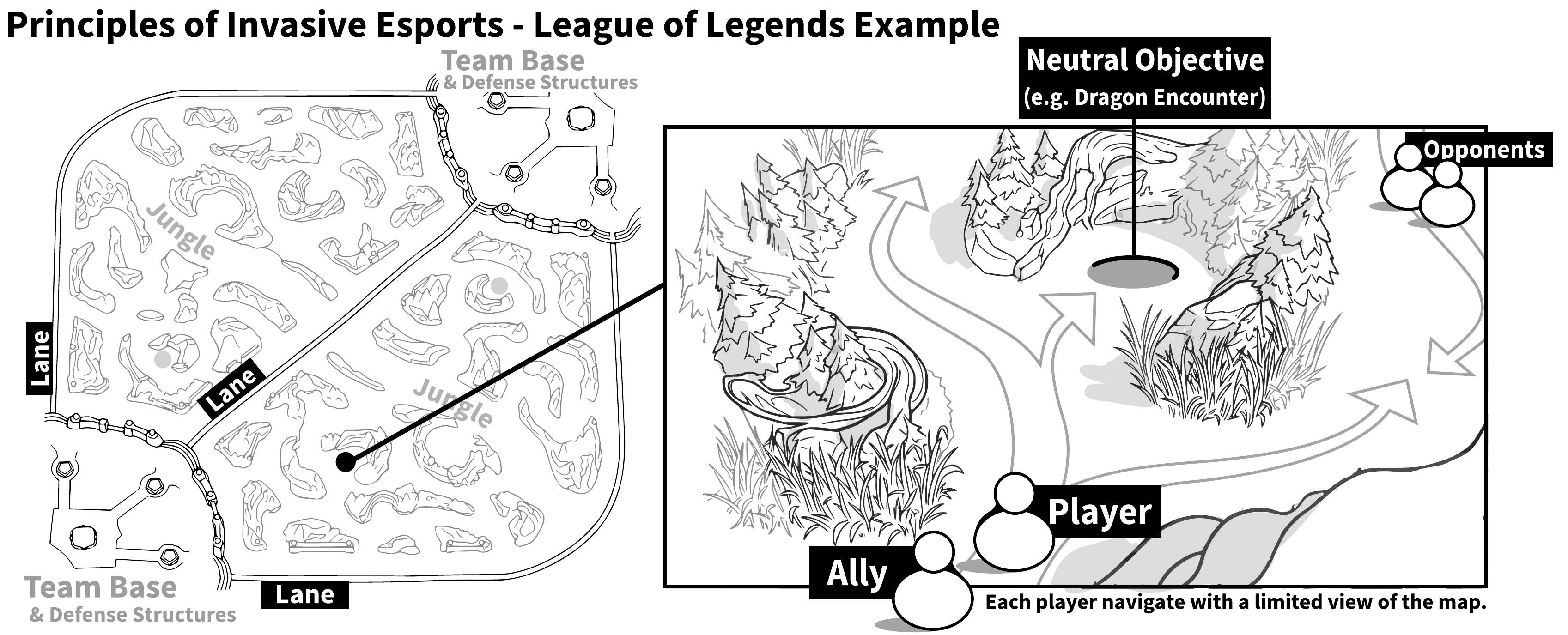}
  \Description{The figure presents a simplified battle arena minimap adapted from League of Legends, highlighting the three lanes, jungle and each team's base on the left. A specific encounter in the jungle near a dragon is highlighted on the right, illustrating how players must choose where to navigate and the possibility of encountering opponents and allies along the way. }
  \caption{\edit{(Left) A typical League of Legends battle arena places each team's base in opposite corners, requiring players to navigate in either lanes or through jungle, where additional neutral objectives exists. (Right) A simplified example of players' view of the map, as they approach a neutral dragon encounter objective.}}
  \label{fig:league-of-legends-example}
\vspace*{-10pt}
\end{figure}

To clarify how the proposed EPS framework can be applied in practice, the following section presents two complementary examples drawn from a single completely hypothetical team, who is aiming to improve their in-game rank in League of Legends, the most widely studied esports titles in performance, coaching, teamwork, and communication research~\cite{bubnaEsportsCoachingApplying2024,novakPerformanceAnalysisEsports2020, tanCommunicationSequencesIndicate2022}. It is a five-versus-five Multiplayer Online Battle Arena game in which each team controls unique player avatars, known as champions, and attempts to destroy the opposing team’s central base structure, the Nexus. The standard map contains three lanes, jungle areas between lanes, defensive structures such as towers and inhibitors, and neutral objectives such as Dragon (see Fig.~\ref{fig:league-of-legends-example}), Rift Herald, and Baron Nashor. Players gain gold and experience by defeating minions, opposing champions, structures, and neutral objectives, allowing their champions to become stronger over time. Because players are distributed across the map with limited vision (similar to Fig.~\ref{fig:league-of-legends-example}), teams must coordinate information, positioning, timing, and objective control in order to create advantages and prevent ambushes. These features make League of Legends a useful example of an invasion-based esport in which teams continually contest space, resources, and strategic objectives.

\subsection{The Background Story} 

Marcus is the coach of a semi-professional League of Legends team attempting to reach Challenger tier (i.e., the highest possible rank category in League of Legends) in preparation for an upcoming trial with a professional organisation. The team includes five highly motivated players who have developed most of their skills through ranked play, amateur tournaments, online guides, VOD reviews, and informal feedback from other players. Each player is individually capable, and several have reached high ranks on their own. However, as a team, their performances remain inconsistent. At times, they look organised and confident, moving together to secure objectives and pressure the map. At other times, the same players appear disconnected, with small disagreements during play quickly turning into lost tempo, poor positioning, and avoidable deaths.

One of Marcus’s main challenges is that the team often disagrees about the source of its problems. After a loss, the players usually identify the same key moment as decisive, but they rarely agree on what went wrong. Kai, the team’s jungler, often argues that fights are lost because players fail to execute mechanically, such as missing key abilities, mis-timing engages, or failing to use summoner spells at the right moment. Elias, the mid laner, tends to see the problem differently. From his perspective, the team’s main weakness is decision-making, particularly around whether to contest objectives, continue fighting, or disengage and reset. Sofia, the support player, frequently points to communication and positioning, arguing that players make reasonable individual choices but fail to act from the same tactical understanding.

These disagreements become most visible around early- to mid-game dragon contests. In several recent matches, the team successfully secured dragon(s) but failed to convert the objective into a lasting advantage. In one game, two players immediately pushed forward into enemy territory after the dragon was taken, while the remaining players recalled to spend gold and reset vision. In another, the team hesitated after winning the objective (dragons), allowing the opponent to regroup and regain control of the river. For Marcus, these moments suggest that the problem is not simply whether the team wins or loses dragon(s). Rather, the deeper issue is how the players interpret the situation immediately before and after the objective is taken.

\looseness-1   To address this, Marcus introduces the EPS framework into the team’s review process. His aim is not to prove that one player’s interpretation is correct, but to give the team a shared language for analysing performance. Instead of labelling a failed sequence as “bad mechanics” or “bad decision-making”, the team begins to ask whether the breakdown occurred at the level of strategy, tactics, tasks, actions, or operations. The first example shows how Marcus uses the framework retrospectively during post-game review to analyse a dragon sequence that initially appeared successful but ultimately destabilised the team. The second example shows how the same team later uses this shared understanding during live play to regulate its decision-making in a similar situation. 
 
\subsection{Example Scenario 1: Diagnostic Screening}

In one of the reviewed matches, Marcus and the team identified a mid-game dragon contest as the moment where the rapport of strength shifted decisively. At first, the play appeared successful. The team secured dragon, gained the associated team-wide buff (i.e., a game mechanic which provides the team an advantage), and seemed to have created an opportunity to extend control over the map. However, the sequence that followed produced the opposite effect. The team lost tempo, became spatially fragmented, and gradually conceded map control back to the opponent.

\looseness-1  The initial review quickly revealed the same disagreement that had characterised the team’s recent post-game discussions. Kai argued that the play broke down because of poor execution, particularly the timing and use of champion abilities during and after the fight. Elias suggested that the problem was decision-making, especially the choice to keep pressuring after the objective had been secured. Sofia, meanwhile, pointed to a lack of shared communication about whether the team should reset or continue forward. Rather than resolving this disagreement through opinion, Marcus used the EPS framework to guide the review. The team deliberately moved through the hierarchy of strategy, tactics, tasks, actions, and operations to determine where instability first emerged.

At the strategic level, the team clarified their overarching direction at that stage of the match. Their aim was to build economic advantages during the laning phase through efficient farming, wave management, and resource collection, then convert those advantages into item progression and objective control during the mid-game. Dragon was therefore not pursued as an isolated gamble, but as part of a broader superiority sequence. Securing it would provide a team-wide buff and allow the team to consolidate the advantages gained earlier in the match, provided that sufficient stability was maintained after the fight. Upon review, the team concluded that contesting dragon was strategically sound. The objective aligned with their longer-term aim of converting early economic advantages into contextual and mechanical superiorities. No fundamental flaw was identified at the strategic layer.

Attention then shifted to the tactical layer, where Marcus located the critical destabilising moment. The review showed that instability did not originate during the dragon contest itself, but immediately after its success. Having secured the objective, the team temporarily held contextual superiority and entered a defence–attack transition. At this point, they had to decide whether to consolidate the gains they had already made or extend pressure into the opponent’s territory. Under such conditions, a principle such as depth or cover would have prioritised the protection of damaged teammates, restoration of vision through wards, and preparation for a coordinated offensive transition once the team had regained structure. 

Instead, the players enacted conflicting tactical principles. Two players advanced aggressively into enemy territory, interpreting the dragon capture as a cue to enact penetration and extend the advantage immediately. At the same time, other players disengaged and recalled to base, implicitly adopting a stabilising logic closer to control and restraint. The absence of a shared principle of play resulted in spatial fragmentation. The team’s temporary advantage was lost, and the isolated players became vulnerable to counter-engagement. Instability was therefore reintroduced not through a mechanical failure, but through a breakdown in tactical coherence during the transition phase.

The team then examined the task layer. The immediate task after securing dragon was not simply to “advance” or “retreat”, but to consolidate the gained superiorities while preserving structural stability. This required the team to decide how their newly acquired contextual and mechanical advantages could be stabilised under conditions of partial health depletion, reduced vision control, and temporary positional advantage. Upon review, the team agreed that this task was appropriate. The problem was not that the players misunderstood the broad task of consolidation. Rather, the problem was that they translated this task into different tactical responses.

At the action layer, the review focused on the specific goal-directed behaviours enacted after dragon was secured. The team examined which players crossed into enemy territory, who began pressuring the nearest tower, who initiated recall, and whether major abilities, including ultimates, were used in ways that aligned with the team’s intended task. When considered in isolation, these actions were not obviously irrational. The players who advanced appeared to interpret consolidation as immediate territorial extension, while the players who recalled interpreted consolidation as stabilisation and reset. In both cases, the actions made local sense. However, they were not coordinated by a shared tactical principle.

Operational execution was then examined as a distinct layer. This analysis focused on fine-grained perceptual–motor performance, including skill-shot accuracy, champion control, and micro-positioning during the actions that followed dragon. Missed abilities and imperfect micro-positioning were observed across several players. However, these slips were similar to those that had occurred earlier in the match without producing a decisive shift in advantage. No concentrated mechanical breakdown was identified at the moment instability emerged.

By comparing the layers of the hierarchy, Marcus and the team concluded that the weakest structural link was located at the tactical layer. The strategic decision to contest dragon was coherent, the immediate task of consolidation was appropriate, and the mechanical imperfections were not decisive. Rather, instability emerged because players enacted conflicting Principles of Play during the defence–attack transition after the dragon encounter. Some players attempted to extend pressure through penetration, while others attempted to stabilise through reset and restraint. This simultaneous enactment of incompatible tactical principles produced loss of cohesion, premature extension, and vulnerability to counter-engagement.

This example illustrates how diagnostic screening can help teams move beyond post-game disagreement and identify where a performance breakdown most likely originated. Without the EPS framework, the same sequence could plausibly be interpreted as a mechanical failure, a poor decision, or a communication problem. By traversing the hierarchy, the team was able to distinguish between slips at the operational level, locally coherent actions at the action layer, and the deeper tactical incoherence that made those actions collectively unstable. The review therefore redirected training away from general claims about poor execution and toward clearer communication, shared tactical language, and collective selection of Principles of Play during transition phases.

\subsection{Example Scenario 2: Real-Time Screening}

In a subsequent training match, a similar mid-game dragon situation emerged. Marcus’s team successfully secured the objective, but the exchange came at a cost: two players were eliminated before the fight concluded. The team had gained the dragon buff, but they had not gained numerical superiority. In the previous match, this was precisely the kind of situation in which instability had emerged during the defence–attack transition. This time, however, the players approached the post-dragon phase with greater awareness of the need for tactical coherence.

Following the earlier diagnostic review, Marcus had encouraged the team to use brief real-time screening after major in-game events. The aim was not to move through the full EPS hierarchy during live play, but to pause long enough to establish a shared tactical orientation. Immediately after dragon was secured, a brief lull occurred. The opponent did not initiate an immediate counter-engagement, and the surviving players had a short window to regulate their next move. The team quickly communicated the relevant state of play: dragon secured, two teammates down, limited vision beyond the river, and no immediate basis for a safe forward push. Although the dragon buff provided a mechanical advantage, the team recognised that this advantage could be lost if the surviving players again split between incompatible tactical principles.

In light of the previous review, the team anchored its immediate decision to Offensive Unity and Balance. Offensive Unity required the surviving players to maintain structural cohesion, avoid spatial dispersion, and preserve collective integrity while the team was incomplete. Balance required them to maintain defensive coverage across vulnerable zones rather than over-commit to forward pressure. Together, these principles helped the team protect against counter-engagement while preserving the mechanical and positional advantages gained from the dragon fight until the eliminated players could re-spawn.

These principles were communicated succinctly and functioned as shared regulatory anchors. The surviving players withdrew to defensible positions, restored vision control where possible, and avoided advancing into enemy territory despite the presence of the dragon buff. No attempt was made to pressure towers or force an additional engagement while re-spawn timers were active. The emphasis was not on maximising immediate gain, but on preventing self-generated instability.

Once the two eliminated players re-spawned and rejoined the team, the tactical situation changed. The team now had the numbers, structure, and dragon buff required to apply pressure more safely. At this point, they collectively transitioned from Balance and Offensive Unity to Penetration, advancing into contested space as a coordinated group. Unlike the previous match, no divergence in tactical principle selection occurred. The shift from stabilising principles to an offensive principle was sequenced deliberately and enacted coherently.

In this instance, real-time screening operated as tactical-layer regulation under time pressure. The team did not conduct a full diagnostic traversal of the EPS hierarchy. Instead, shared PoP terminology enabled them to compress complex situational information into a rapid collective decision. By doing so, they preserved the superiorities gained through the dragon fight and avoided recreating the self-generated instability that had undermined their performance in the earlier match.

\section{Implications for Coaching, Analysis and Performance}

The EPS framework has several proposed implications for coaching practice, analytic tool development, and team communication. These implications should be understood as conceptual and applied possibilities rather than empirically validated outcomes. The examples above illustrate how the framework may support more systematic review and decision-making, but future work is required to test whether its use improves coaching effectiveness, player learning, or team performance in real-world esports environments.

For coaching practice, the EPS framework may provide a structure for organising VOD review and avoiding unproductive micro-critique. Rather than treating every failed sequence as a collection of individual mistakes, coaches can use the hierarchy to distinguish whether a breakdown originated at the level of strategy, tactics, tasks, actions, or operations. This distinction is important because different errors require different interventions. A strategic problem may require changes to match preparation or win-condition planning; a tactical problem may require clearer selection and communication of Principles of Play; a task-level problem may require players to clarify the immediate objective; an action-level problem may require work on situational judgement; and an operational problem may require targeted mechanical practice. In this way, the EPS framework may help coaches to better identify the most important error to target for the most effective training. As such this could reduce the amount of time spent on VOD reviews and improve the speed in which performance can be improved.

For analytic and digital tools, the EPS framework may inform the design of replay systems and visual dashboards. Gameplay events could be annotated according to their level in the hierarchy, the principles of play involved, and the superiorities gained, lost, preserved, or denied. For example, a replay tool might tag whether a failed objective contest involved a tactical mismatch between intended and enacted PoP, a task-level ambiguity about whether to reset or continue pressure, or an operational slip such as a missed input or poorly executed ability. Aggregated across matches, these annotations could allow dashboards to show where breakdowns cluster, such as repeated tactical incoherence during attack-defence transitions, recurrent operational slips during high-pressure fights, or systematic failure to convert informational superiority into positional control. Such tools would not replace coach interpretation, but could help direct attention toward recurring patterns that are difficult to identify through unstructured review.

For team communication, the EPS framework may offer a shared vocabulary that reduces cross-talk during both review and live play. In many team environments, players may disagree because they are analysing the same event at different levels of the EPS framework. One player may describe a failed play as poor mechanics, another as a bad decision, and another as a communication problem. The EPS framework provides a way to make these distinctions explicit. A coach or player can state, for example, “this is a task-level issue, not a mechanics issue”, or “the action made sense individually, but the tactical principle was not shared”. This language may support more precise self-reflection by helping players locate their own contribution within the broader structure of team performance. It may also support real-time coordination by allowing teams to compress complex situational information into shared tactical anchors, such as stabilising through Balance before transitioning collectively to Penetration.

These implications suggest that the EPS framework may function as a bridge between theory and practice. For coaches, it offers a way to structure review and set level-specific learning goals. For analysts and tool developers, it offers a taxonomy for organising gameplay data and feedback. For players and teams, it offers a shared language for reflecting on performance and coordinating decisions under pressure. However, these applications remain proposed uses of the EPS framework.

\section{Future Research}

The integrated EPS framework presented in this paper offers a foundation for multi-level analysis of invasion-based esports, but several areas warrant empirical and applied development. One priority is to examine how PoP and the proposed forms of superiority manifest across different game titles, competitive levels and player roles. Comparative studies could identify, which principles and superiorities are robust across contexts and which require game-specific calibration, thereby testing the claim that the EPS framework is broadly applicable within invasion-based esports.

\looseness-1  A second priority is to evaluate the framework’s pedagogical utility in applied settings. This would first involve examining where practitioners, such as coaches and educators can apply the EPS framework effectively during training, review, and match preparation. Identifying any points of difficulty would help clarify which aspects of the framework require simplification, adaptation, or additional instructional support. Experimental or quasi-experimental studies could then compare training programmes that explicitly use the EPS framework with those that rely on existing informal methods, with outcomes including decision-making processes, team coordination, tactical communication, and self-regulatory behaviour. Qualitative work with coaches and players would complement these designs by examining how practitioners actually use the EPS framework during review–plan–do cycles, and how it affects communication, shared understanding, and role clarity in teams.

Future research should also investigate how players internalise and transition between strategic, tactical, action, and operational routines over time. Longitudinal and event-based methods could be used to track how repeated exposure to the framework shapes attention, attribution patterns, and the granularity of self-reflection, particularly under conditions of stress or high cognitive load. Coupling the framework with multimodal behavioural data sources such as eye tracking, keystroke logs, communication data, replay annotations, or physiological measures may deepen understanding of how specific interaction patterns and superiorities emerge in real time.

\edit{A further priority concerns the development of analytics and training tools based on the EPS framework. Performance analysts and tool developers could use the EPS framework as a taxonomy for organising data and feedback. Strategic calls, tactical choices, task demands, action patterns, and operational events could be tagged according to their level in the hierarchy, the PoP involved, and the superiorities gained or lost. Telemetry systems and replay tools that adopt such a structure could help surface patterns that are otherwise difficult to see, such as systematic mismatches between intended principles and realised superiorities, recurring tactical incoherence during transitions, or operational slips that undermine otherwise sound tactics. Future research could therefore examine whether framework-informed tagging systems improve the efficiency, accuracy, and usefulness of post-game review, both for coaches and for players engaging in self-analysis.}

\looseness-1  Finally, broader adoption and refinement of the framework are likely to benefit from co-design approaches. Iterative development involving players, coaches, analysts, educators, and tool developers could be used to adapt terminology, visualisations, tagging systems, and training materials for different competitive and educational contexts. Such work would help ensure that the EPS framework remains usable in under-resourced environments while still supporting the more sophisticated analytic needs of professional teams, performance analysts, and esports technology developers.

\section{Conclusion}

This paper has presented a conceptual integrated framework for analysing gameplay in invasion-based esports that links strategic intentions to tactical choices, concrete in-game actions and fine-grained mechanical execution. By integrating the Principles of Play and Core Task Framework to the context of team-based digital competition and a structured account of superiorities and the stability instability cycle, the EPS framework offers a coherent and comprehensive hierarchy for describing how advantages are created, maintained and lost during play. It's potential application in practice has been illustrated through the two hypothetical case examples, the next step is to apply the EPS framework in a real gaming team environment. 

The EPS framework contributes both analytically and pedagogically to the development of coaching practice in esports and targeted player development specifically with focus on strategy and tactical decision-making. Analytically, it provides a set of interconnected constructs that can be used to decompose performance breakdowns across levels, distinguishing failures of strategic logic, tactical principle selection, action-level judgement and operational execution. Pedagogically, it supports the design of coaching and learning environments in which players review, plan and act using a shared structure, translating broad strategic goals into trainable behaviours and making self-regulatory processes more explicit. The hypothetical applications demonstrate how the EPS framework can inform clip review, practice planning and self-reflection in professional, grassroots and scholastic settings.

\looseness-1  Although the EPS framework has been developed with invasion-based esports as its primary target, the underlying logic of hierarchical decision-making, PoP and superiorities may have relevance for other high-tempo digital environments where teams coordinate under uncertainty. Future work should test the EPS framework's applicability across genres, evaluate its impact in real-world coaching and educational interventions and explore its potential to inform tools that visualise, annotate and explain gameplay data. For now, the EPS framework provides a structured starting point for researchers and practitioners (e.g., coaches and educators) who seek to connect theoretical insight about decision-making and coordination with practical efforts to improve performance and learning in esports.

\bibliographystyle{ACM-Reference-Format}
\bibliography{HTSmodelpaper}

@article{abbott_perceptions_2023,
  title = {Perceptions of Effective Training Practices in League of Legends: A Qualitative Exploration},
  shorttitle = {Perceptions of Effective Training Practices in League of Legends},
  author = {Abbott, Callum and Watson, Matthew and Birch, Phil},
  year = 2023,
  month = jan,
  journal = {Journal of Electronic Gaming and Esports},
  volume = {1},
  number = {1},
  pages = {jege.2022--0011},
  issn = {2836-3523},
  doi = {10.1123/jege.2022-0011},
  urldate = {2026-01-19},
  langid = {english}
}

@article{abraham_taking_2011,
  title = {Taking the next Step: {{Ways}} Forward for Coaching Science},
  shorttitle = {Taking the next Step},
  author = {Abraham, Andrew and Collins, Dave},
  year = 2011,
  month = nov,
  journal = {Quest (Grand Rapids, Mich.)},
  volume = {63},
  number = {4},
  pages = {366--384},
  issn = {0033-6297, 1543-2750},
  doi = {10.1080/00336297.2011.10483687},
  urldate = {2025-07-09},
  langid = {english}
}

@article{buekers_strategy_2020,
  title = {Strategy and Tactics in Sports from an Ecological-Dynamical-Perspective: {{What}} Is in There for Coaches and Players?},
  shorttitle = {Strategy and Tactics in Sports from an Ecological-Dynamical-Perspective},
  author = {Buekers, Martinus and Montagne, Gilles and {Ib{\'a}{\~n}ez-Gij{\'o}n}, Jorge},
  year = 2020,
  journal = {Movement \& Sport Sciences - Science \& Motricit\'e},
  volume = {106},
  number = {108},
  pages = {1--11},
  issn = {2118-5735, 2118-5743},
  doi = {10.1051/sm/2019026},
  urldate = {2025-06-04},
  langid = {english}
}

@article{clemente_systemic_2014,
  title = {A Systemic Overview of Football Game: {{The}} Principles behind the Game},
  shorttitle = {A Systemic Overview of Football Game},
  author = {Clemente, Filipe Manuel and {Martins, Fernando Manuel Louren\c co}},
  year = 2014,
  journal = {Journal of Human Sport and Exercise},
  volume = {9},
  number = {2},
  pages = {656--667},
  issn = {19885202},
  doi = {10.14198/jhse.2014.92.05},
  urldate = {2021-10-26},
  langid = {english}
}

@article{hougaard_aiming_2024,
  title = {Aiming, Pointing, Steering: A Core Task Analysis Framework for Gameplay},
  shorttitle = {Aiming, Pointing, Steering},
  author = {Hougaard, Bastian Ils{\o} and Knoche, Hendrik},
  year = 2024,
  month = oct,
  journal = {Proceedings of the ACM on Human-Computer Interaction},
  volume = {8},
  number = {CHI PLAY},
  pages = {1--48},
  issn = {2573-0142},
  doi = {10.1145/3677057},
  urldate = {2025-01-14},
  langid = {english}
}

@phdthesis{juntunen_challenges_2022,
  type = {Bachelor's Thesis},
  title = {The Challenges of European Esports Coaching from the Coach's Perspective},
  author = {Juntunen, Tero},
  year = 2022,
  address = {Jyv\"askyl\"a},
  urldate = {2025-06-15},
  langid = {english},
  school = {Jamk University of Applied Sciences}
}

@article{kleinman_trust_2024,
  title = {"{{Trust}} the Process": {{An}} Exploratory Study of Process Visualizations for Self-Reflection in League of Legends},
  shorttitle = {"{{Trust}} the Process"},
  author = {Kleinman, Erica and Xu, Jason and Pfau, Johannes and {Seif El-Nasr}, Magy},
  year = 2024,
  month = oct,
  journal = {Proceedings of the ACM on Human-Computer Interaction},
  volume = {8},
  number = {CHI PLAY},
  pages = {1--28},
  issn = {2573-0142},
  doi = {10.1145/3677111},
  urldate = {2025-02-12},
  langid = {english}
}

@inproceedings{lee_crafting_2025,
  title = {Crafting Champions: {{An}} Observation Study of Esports Coaching Processes},
  shorttitle = {Crafting Champions},
  booktitle = {Proceedings of the 2025 {{CHI}} Conference on Human Factors in Computing Systems},
  author = {Lee, Hanbyeol and Kleinman, Erica and Kim, Namsub and Park, Sangbeom and Harteveld, Casper and Lee, Byungjoo},
  year = 2025,
  month = apr,
  pages = {1--20},
  publisher = {ACM},
  address = {Yokohama Japan},
  doi = {10.1145/3706598.3713141},
  urldate = {2025-06-15},
  eventtitle = {{{CHI}} 2025: {{CHI}} Conference on Human Factors in Computing Systems},
  isbn = {979-8-4007-1394-1},
  langid = {english}
}

@techreport{newzoo_newzoo_2022,
  title = {Newzoo Free Global Esports Live Streaming Market Report},
  institution ={Newzoo},
  author = {{Newzoo}},
  year = 2022
}

@book{norman_design_2013,
  title = {The Design of Everyday Things},
  author = {Norman, Don},
  year = 2013,
  edition = {Revised and Expanded Edition},
  publisher = {Basic Books},
  address = {New York},
  isbn = {978-0-465-07299-6},
  langid = {english}
}

@article{palmer_cooperative_2023,
  title = {Cooperative Networks in Team Invasion Games: {{A}} Systematic Mapping Review},
  shorttitle = {Cooperative Networks in Team Invasion Games},
  author = {Palmer, Sam and Novak, Andrew R. and Tribolet, Rhys and Watsford, Mark L. and Fransen, Job},
  year = 2023,
  month = dec,
  journal = {International Journal of Sports Science \& Coaching},
  volume = {18},
  number = {6},
  pages = {2347--2359},
  issn = {1747-9541, 2048-397X},
  doi = {10.1177/17479541231177133},
  urldate = {2025-07-08},
  langid = {english}
}

@article{sabtan_current_2022,
  title = {Current Practice and Challenges in Coaching {{Esports}} Players: {{An}} Interview Study with League of Legends Professional Team Coaches},
  shorttitle = {Current Practice and Challenges in Coaching {{Esports}} Players},
  author = {Sabtan, Bader and Cao, Shi and Paul, Naomi},
  year = 2022,
  month = may,
  journal = {Entertainment Computing},
  volume = {42},
  pages = {100481},
  issn = {18759521},
  doi = {10.1016/j.entcom.2022.100481},
  urldate = {2025-03-07},
  langid = {english}
}

@article{teoldo_da_costa_tactical_2009,
  title = {Tactical Principles of Soccer: Concepts and Application},
  author = {{Teoldo da Costa}, Israel and {Garganta da Silva}, J{\'u}lio Manuel and Greco, Pablo Juan and Mesquita, Isabel},
  year = 2009,
  journal = {Motriz : revista de educa\c c\~ao f\'isica. UNESP},
  volume = {15},
  number = {3},
  pages = {657--668},
  langid = {english}
}

@article{watson_i_2025,
  title = {``{{I}} Don't Believe Any Qualifications Are Required'': {{Exploring}} Global Stakeholders' Perspectives towards the Developmental Experiences of Esports Coaches},
  shorttitle = {``{{I}} Don't Believe Any Qualifications Are Required''},
  author = {Watson, Matthew and Trotter, Michael G. and Laborde, Sylvain and Leeder, Thomas M.},
  year = 2025,
  month = jul,
  journal = {Education Sciences},
  volume = {15},
  number = {7},
  pages = {858},
  issn = {2227-7102},
  doi = {10.3390/educsci15070858},
  urldate = {2026-02-18},
  langid = {english}
}

@article{watson_introducing_2022,
  title = {Introducing Esports Coaching to Sport Coaching (Not as Sport Coaching)},
  author = {Watson, Matthew and Smith, David and Fenton, Jack and {Pedraza-Ramirez}, Ismael and Laborde, Sylvain and Cronin, Colum},
  year = 2022,
  month = sep,
  volume = 14,
  issue = 2,
  journal = {Sports Coaching Review},
  pages = {1--20},
  issn = {2164-0629, 2164-0637},
  doi = {10.1080/21640629.2022.2123960},
  urldate = {2025-03-07},
  langid = {english}
}

@phdthesis{wharton_expertise_2014,
  title = {Expertise in Coaching Interceptive Sports: {{A}} Grounded Theory Model},
  shorttitle = {Expertise in Coaching Interceptive Sports},
  author = {Wharton, Donald Lee},
  year = 2014,
  month = nov,
  doi = {10.14264/uql.2015.106},
  urldate = {2019-10-14},
  langid = {english},
  school = {The University of Queensland}
}

@article{TrotterGoals2025,
  title = {Goal Systems and Regulatory Processes in Esports: A Pilot Study on League of Legends Players},
  author = {Trotter, Michael and Svensson, Joar and Tan, Evelyn},
  year = 2025,
  journal = {Journal of Electronic Gaming and Esports},
  volume = {3},
  number = {1},
  pages = {jege.2024-0048},
  publisher = {Human Kinetics},
  address = {Champaign IL, USA},
  doi = {10.1123/jege.2024-0048}
}

@article{bednySystemicStructuralTheoryActivity2005,
  title = {The {{Systemic-Structural Theory}} of {{Activity}}: {{Applications}} to the {{Study}} of {{Human Work}}},
  shorttitle = {The {{Systemic-Structural Theory}} of {{Activity}}},
  author = {Bedny, Gregory Z. and Harris, Steven Robert},
  year = 2005,
  month = may,
  journal = {Mind, Culture, and Activity},
  volume = {12},
  number = {2},
  pages = {128--147},
  issn = {1074-9039, 1532-7884},
  doi = {10.1207/s15327884mca1202_4},
  urldate = {2025-02-14},
  langid = {english}
}

@article{bubnaEsportsCoachingApplying2024,
  title = {Esports {{Coaching}}: {{Applying}} a {{Constraints-Led Approach}} to {{Develop Team Coordination}} and {{Communication}}},
  shorttitle = {Esports {{Coaching}}},
  author = {Bubna, Kabir and Trotter, Michael G. and Watson, Matthew},
  year = 2024,
  month = jan,
  journal = {Case Studies in Sport and Exercise Psychology},
  volume = {8},
  number = {S1},
  pages = {S1-9-S1-17},
  issn = {2470-4849, 2470-4857},
  doi = {10.1123/cssep.2023-0022},
  urldate = {2024-03-02},
  langid = {english}
}

@article{innsDecisionmakingAssessmentsYouth2023,
  title = {Decision-Making Assessments in Youth Team Invasion Game Athletes: {{A}} Systematic Scoping Review},
  shorttitle = {Decision-Making Assessments in Youth Team Invasion Game Athletes},
  author = {Inns, Joshua and Petancevski, Emma L and Novak, Andrew R and Fransen, Job},
  year = 2023,
  month = dec,
  journal = {International Journal of Sports Science \& Coaching},
  volume = {18},
  number = {6},
  pages = {2360--2381},
  issn = {1747-9541, 2048-397X},
  doi = {10.1177/17479541231185779},
  urldate = {2026-05-22},
  langid = {english}
}

@article{nagorskyStructurePerformanceTraining2020,
  title = {The Structure of Performance and Training in Esports},
  author = {Nagorsky, Eugen and Wiemeyer, Josef},
  editor = {Pess\^oa Filho, Dalton M\"uller},
  year = 2020,
  month = aug,
  journal = {PLOS ONE},
  volume = {15},
  number = {8},
  pages = {e0237584},
  issn = {1932-6203},
  doi = {10.1371/journal.pone.0237584},
  urldate = {2026-05-22},
  langid = {english}
}

@article{novakPerformanceAnalysisEsports2020,
  title = {Performance Analysis in Esports: Modelling Performance at the 2018 {{League}} of {{Legends World Championship}}},
  shorttitle = {Performance Analysis in Esports},
  author = {Novak, Andrew R and Bennett, Kyle Jm and Pluss, Matthew A and Fransen, Job},
  year = 2020,
  month = dec,
  journal = {International Journal of Sports Science \& Coaching},
  volume = {15},
  number = {5-6},
  pages = {809--817},
  issn = {1747-9541, 2048-397X},
  doi = {10.1177/1747954120932853},
  urldate = {2026-05-22},
  langid = {english}
}

@article{rogersKovaaKsAimTrainer2024,
  title = {{{KovaaK}}'s Aim Trainer as a Reliable Metrics Platform for Assessing Shooting Proficiency in Esports Players: A Pilot Study},
  shorttitle = {{{KovaaK}}'s Aim Trainer as a Reliable Metrics Platform for Assessing Shooting Proficiency in Esports Players},
  author = {Rogers, Ethan J. and Trotter, Michael G. and Johnson, Daniel and Desbrow, Ben and King, Neil},
  year = 2024,
  month = feb,
  journal = {Frontiers in Sports and Active Living},
  volume = {6},
  pages = {1309991},
  issn = {2624-9367},
  doi = {10.3389/fspor.2024.1309991},
  urldate = {2024-05-03},
  langid = {english}
}

@article{rumpfManipulatingPlayerNumbers2026,
  title = {Manipulating Player Numbers in Small-Sided Soccer Games: Effects on Numerical Superiority and Inferiority across Physiological, Physical, Technical, and Tactical Performance},
  shorttitle = {Manipulating Player Numbers in Small-Sided Soccer Games},
  author = {Rumpf, Michael C. and J\"ager, Johannes and Lloyd, Rhodri S. and Lochmann, Matthias},
  year = 2026,
  month = apr,
  journal = {Frontiers in Sports and Active Living},
  volume = {8},
  pages = {1813770},
  issn = {2624-9367},
  doi = {10.3389/fspor.2026.1813770},
  urldate = {2026-05-22},
  langid = {english}
}

@article{sharpeIndexingEsportPerformance2023,
  title = {Indexing {{Esport Performance}}},
  author = {Sharpe, Benjamin T. and Besombes, Nicolas and Welsh, Matthew R. and Birch, Phil D.J.},
  year = 2023,
  month = jan,
  journal = {Journal of Electronic Gaming and Esports},
  volume = {1},
  number = {1},
  pages = {jege.2022-0017},
  issn = {2836-3523},
  doi = {10.1123/jege.2022-0017},
  urldate = {2023-03-13},
  langid = {english}
}

@article{tanCommunicationSequencesIndicate2022,
  title = {Communication {{Sequences Indicate Team Cohesion}}: {{A Mixed-Methods Study}} of {{Ad Hoc League}} of {{Legends Teams}}},
  shorttitle = {Communication {{Sequences Indicate Team Cohesion}}},
  author = {Tan, Evelyn T S and Rogers, Katja and Nacke, Lennart E. and Drachen, Anders and Wade, Alex},
  year = 2022,
  month = oct,
  journal = {Proceedings of the ACM on Human-Computer Interaction},
  volume = {6},
  number = {CHI PLAY},
  pages = {1--27},
  issn = {2573-0142},
  doi = {10.1145/3549488},
  urldate = {2024-06-05},
  langid = {english}
}

@article{vilarVaryingNumbersPlayers2014,
  title = {Varying {{Numbers}} of {{Players}} in {{Small-Sided Soccer Games Modifies Action Opportunities}} during {{Training}}},
  author = {Vilar, Lu\'is and Esteves, Pedro T. and Travassos, Bruno and Passos, Pedro and {Lago-Pe\~nas}, Carlos and Davids, Keith},
  year = 2014,
  month = oct,
  journal = {International Journal of Sports Science \& Coaching},
  volume = {9},
  number = {5},
  pages = {1007--1018},
  issn = {1747-9541, 2048-397X},
  doi = {10.1260/1747-9541.9.5.1007},
  urldate = {2026-05-22},
  langid = {english}
}

@article{xiaWhatContributesSuccess2019,
  title = {What {{Contributes}} to {{Success}} in {{MOBA Games}}? {{An Empirical Study}} of {{Defense}} of the {{Ancients}} 2},
  shorttitle = {What {{Contributes}} to {{Success}} in {{MOBA Games}}?},
  author = {Xia, Bang and Wang, Huiwen and Zhou, Ronggang},
  year = 2019,
  month = jul,
  journal = {Games and Culture},
  volume = {14},
  number = {5},
  pages = {498--522},
  issn = {1555-4120, 1555-4139},
  doi = {10.1177/1555412017710599},
  urldate = {2026-05-22},
  langid = {english}
}

@mastersthesis{Jalovaara2024WinProbEsports,
  author       = {Jalovaara, Perttu},
  title        = {Win Probability Estimation for Strategic Decision-Making in Esports},
  school       = {Aalto University},
  year         = {2024},
  type         = {Master's thesis},
  url          = {https://sal.aalto.fi/publications/pdf-files/theses/mas/tjal24a_public.pdf}
}

@article{Memmert2008GPAIConceptual,
  author       = {Memmert, Daniel and Harvey, Stephen},
  title        = {The Game Performance Assessment Instrument (GPAI): Some Conceptual and Methodological Issues to Improve the Quality of Game Performance Assessment},
  journal      = {Journal of Teaching in Physical Education},
  year         = {2008},
  volume       = {27},
  number       = {2},
  pages        = {220--240}
}

@misc{Zhang2024SIDOModel,
  author       = {Zhang, Amy X. and Naidu, Pavan},
  title        = {The SIDO Performance Model for League of Legends},
  year         = {2024},
  eprint       = {2403.04873},
  archivePrefix= {arXiv},
  primaryClass = {cs.LG},
  note         = {Preprint}
}

@article{Lord17102020,
author = {Felicity Lord and David B Pyne and Marijke Welvaert and Jocelyn K Mara},
title = {Methods of performance analysis in team invasion sports: A systematic review},
journal = {Journal of Sports Sciences},
volume = {38},
number = {20},
pages = {2338--2349},
year = {2020},
publisher = {Routledge},
doi = {10.1080/02640414.2020.1785185}
}

@article{Oslin1998GPAI,
  author       = {Oslin, Judith L. and Mitchell, Stephen A. and Griffin, Linda L.},
  title        = {The Game Performance Assessment Instrument (GPAI): Development and Preliminary Validation},
  journal      = {Journal of Teaching in Physical Education},
  year         = {1998},
  volume       = {17},
  number       = {2},
  pages        = {231--243},
  doi = {10.1123/jtpe.17.2.231}
}

@incollection{trotter2026decision,
  title={Decision-Making and Strategic Thinking in Esports},
  author={Trotter, Michael and Gredin, Viktor and Kleinmann, Erica and Wadenholt, Gustaf and Svennson, Joar},
  booktitle={The Psychology of Esports Performance},
  pages={75--95},
  year={2026},
  publisher={Routledge},
  address={London}
  
}

\end{document}